\documentclass[10pt]{iopart}

\usepackage{graphicx}
\usepackage{multirow}
\usepackage{cite}
\usepackage{citesort}

\begin{document}

\title{Optimizing Vetoes for Gravitational-wave Transient Searches}

\author{R Essick,$^1$ L Blackburn$^2$ and E Katsavounidis$^3$}
\address{$^1$ MIT, LIGO Laboratory, NW22-295, 185 Albany Str, Cambridge, MA 02139}
\address{$^2$ NASA-GSFC 663, 8800 Greenbelt Rd, Greenbelt, MD 20771}
\address{$^3$ MIT, LIGO Laboratory, NW22-295, 185 Albany Str, Cambridge, MA 02139}
\eads{$^1$ ressick@mit.edu, $^2$ lindy.l.blackburn@nasa.gov, $^3$ kats@mit.edu}

\begin{abstract}

Interferometric gravitational-wave detectors like LIGO, GEO600 and Virgo record a surplus of
information above and beyond possible gravitational-wave events.
These auxiliary channels capture information about
the state of the detector and its surroundings which can be used to
infer potential terrestrial noise sources of some gravitational-wave-like
events.  We present an algorithm addressing the ordering (or equivalently
optimizing) of such information from auxiliary systems
in gravitational-wave
detectors to establish veto conditions in searches
for gravitational-wave
transients. The procedure was used to identify vetoes for searches for
unmodeled
transients by the LIGO and Virgo collaborations during their science runs from
2005 through 2007. In this
work we present
the details of the algorithm; we also use a limited amount of data
from LIGO's past runs
in order to examine the method, compare it with other
methods, and identify
its potential to characterize the instruments themselves. We examine
the dependence
of Receiver Operating Characteristic curves on the various parameters
of the veto method
and the implementation on real data.
We find that the method robustly determines 
important
auxiliary channels, ordering them by the apparent strength of their
correlations to the gravitational-wave channel. This list can substantially
reduce the background of noise events in the gravitational-wave data.
In this way it can identify the source of glitches in the detector
as well as assist in establishing confidence in the detection of
gravitational-wave transients.

\end{abstract}



\maketitle

\section{Introduction}\label{Introduction}

The Laser Interferometer Gravitational-wave Observatory
(LIGO)~\cite{LIGOinstrument}
together with Virgo~\cite{Virgoinstrument}
and GEO600~\cite{GEOinstrument}
form a network of detectors
employing kilometer-scale interferometers to search for
gravitational waves (GWs) from astrophysical and cosmological sources.
One such class of sources is expected to result in short-lived signals
lasting from milliseconds to several seconds within the sensitive
frequency band of the instruments. They may correspond to core-collapse supernovae, neutron star glitches, cosmic string cusps
and kinks, magnetars or some binary compact systems (made up of neutron
stars/black holes)~\cite{Cutler}. Environmental and instrumental noise
sources may generate similar short-lived signals through their
coupling to the GW sensing channel. These signals are
colloquially referred to as ``glitches.'' During the first-generation
instruments' operation (2002-2010), such glitches were non-Gaussian and
non-stationary and presented challenges when
searching for transients of astrophysical origin.
For well modeled GW transient sources (like most of
the binary compact star coalescences), knowledge of the expected
signal waveform significantly helps reject such glitches.
These signal-based vetoes have been developed and  invoked in
existing searches~\cite{CBCS5,CBCS6}.
However, for unmodeled (or poorly modeled) transient
searches, such glitches may drive detection thresholds to significantly
higher values than one would expect for Gaussian
backgrounds~\cite{BurstsS5VSR1,BurstsS6}. 

Interferometric detectors record their physical environment and
detailed interferometry status through thousands of auxiliary channels
that present no or negligible coupling to GWs.
Information from these channels presents an important handle for understanding
(and fixing) the sources of noise in the instruments, reducing the background
and ultimately establishing
confidence in detections. The problem of identifying and ``mechanizing''
the use of information from auxiliary channels is long standing
within the GW data analysis community~\cite{DiCredico:2005,Christensen:2004,Christensen:2005}.

A number of statistical quantities have been developed~\cite{Slutsky:2010} in
order to help characterize the performance of a particular auxiliary channel or
veto strategy, such as {\em veto efficiency}: the
fraction of GW-channel glitches removed, {\em use percentage}: the fraction of
auxiliary channel glitches which can be associated with a GW-channel
glitch~\cite{DiCredico:2005,Isogai:2005},
{\em dead-time}: the effective fraction of analysis
live-time removed when applying the veto strategy, and {\em veto significance}:
the statistical significance of a measured correlation between auxiliary and
GW-channel glitches assuming random coincidence~\cite{Katsavounidis:2006}.
These veto
metrics are most appropriate for a simple veto strategy, such as a coincidence
between the auxiliary and GW-channel glitch within a short specified time
window. Expansions on this approach include making use of our knowledge of
the instrument to anticipate when noise coupling between the auxiliary and GW
channels is strongest and/or consistent with observation~\cite{Ajith1,Ajith2}.
The use of
machine learning algorithms to digest the large amounts of auxiliary channel
information and predict GW-channel glitches is also
an active area of study~\cite{auxmvc}.
Aside from auxiliary information, signal consistency
of the event across multiple instruments, or between the observed signal and
theoretical waveform, provide an extremely powerful way to reject noise
transients. However, accidental transient noise coincidence across multiple
instruments is still a dominant source of background in astrophysical
searches, especially in searches for unmodeled transients.

In this paper we present an algorithm which we will call
{\em Ordered Veto List} (OVL). It exploits the information from the
auxiliary channels in GW detectors and uses a unified ranking metric,
veto efficiency divided by
dead-time (described earlier), in order
to make inferences about the source of GW-channel events recorded at each
detector. The OVL algorithm operates by generating a small time window
surrounding a glitch in an auxiliary channel. If a GW-channel transient is present
within this window, it is assumed to be noise and removed.  An
earlier version of the method, described in~\cite{blackburnthesis}, was used to
identify noise transients in a search for GW bursts during
LIGO's fifth and Virgo's first science runs~\cite{BurstsS5,BurstsS5VSR1}.

OVL addresses several problems with the simple strategy of removing all
live-time associated with a disturbance in an auxiliary channel. First, OVL
identifies only those channels with noise that couples in a statistically
significant way to the GW channel, thus avoiding the unnecessary removal
of live-time.
Second, when multiple auxiliary channels cover the same transient
disturbances in the
instrument, OVL selects the channel which optimally removes background with
minimal loss of live-time. Ultimately, OVL provides an ordered list of rules to
follow for successively removing time from the analysis based on auxiliary channel
information, with the most effective channels at the top of the
list~\cite{Katsavounidis:2006}.
The ordered list also provides a metric representing our confidence that any
particular event is instrumental in origin.

Another hierarchical veto selection method,
{\em h-veto}~\cite{Smith:2011an}, has been developed.
Both methods have a similar
strategy for ranking veto channels. The major
difference is in the figure-of-merit used in the ordering process. The ranking
for {\em h-veto} depends on the statistical significance of the
correlation between the auxiliary and GW channels under the assumption
of a Poisson
process.  This choice favors large statistics, and generally results in a short
list of highly-\emph{efficient} vetoes, where each entry represents the most relevant
physical parameters (e.g. time-scale) of the coupling. OVL, on the other hand,
ranks vetoes by the ratio of \emph{efficiency} divided by the \emph{fractional dead-time}.
This favors
vetoes which have the highest rate of GW-channel transients within their chosen
exclusion windows.  In this ranking, it is common for the same auxiliary channel to
appear multiple times in the list with different thresholds on glitch
strength or different exclusion window durations. Typically a channel is
chosen first at the highest
thresholds (strongest auxiliary channel disturbances) and smallest
exclusion windows,
and only appears later with more relaxed parameters. This choice maximizes
the overall
efficiency obtained using the best veto parameters at a fixed dead-time
threshold at the expense of a longer, interlaced list which can be
more difficult to interpret. Finally, in OVL, the veto windows are explicitly
calculated as a set of non-overlapping time intervals (segments), and
overlaps between
different veto conditions are calculated exactly. This can be important when
considering highly-correlated auxiliary channels with significant overlap.

This paper is organized as follows.
In Section \ref{DescriptionOfTheAlgorithm}, we describe the OVL algorithm,
including construction of veto configurations and their application to GW-channel data.
In Section \ref{ResultsAndAnalysis}, we discuss OVL's performance when applied
to two samples of LIGO data: one month from LIGO's fourth science run (S4) and one
week from LIGO's sixth science run (S6). We evaluate the method's performance using
Receiver Operating Characteristic (ROC) curves and examine some features of the
ordering generated. We conclude in Section \ref{Conclusions}.

\section{Description of the algorithm}\label{DescriptionOfTheAlgorithm}

OVL is based on a simple process, which includes constructing
veto-configurations, the application of those configurations to GW-channel data, and
iteration. 
It uses glitches identified by a generic transient-finding algorithm called
Kleine Welle (KW)~\cite{KW2004}. Each of these glitches is characterized by a
time, duration, frequency content, and significance\footnote{KW significance is
a measure of the likelihood of observing a signal with greater or equal
signal energy assuming a gaussian noise distribution.}. While all our
investigations are based on KW, \emph{any} glitch-finding method able to
provide at least a time and a significance for each event can be used. In this
way, OVL can be adapted to specific searches for different types of GW signals,
or to more general detection problems beyond GW astronomy.

\subsection{Construction of veto configurations}\label{Construction of veto configurations}

\begin{figure}
  \begin{center}
    \includegraphics[width=0.75\textwidth, clip=True, trim=2cm 2.5cm 2cm 3.5cm]{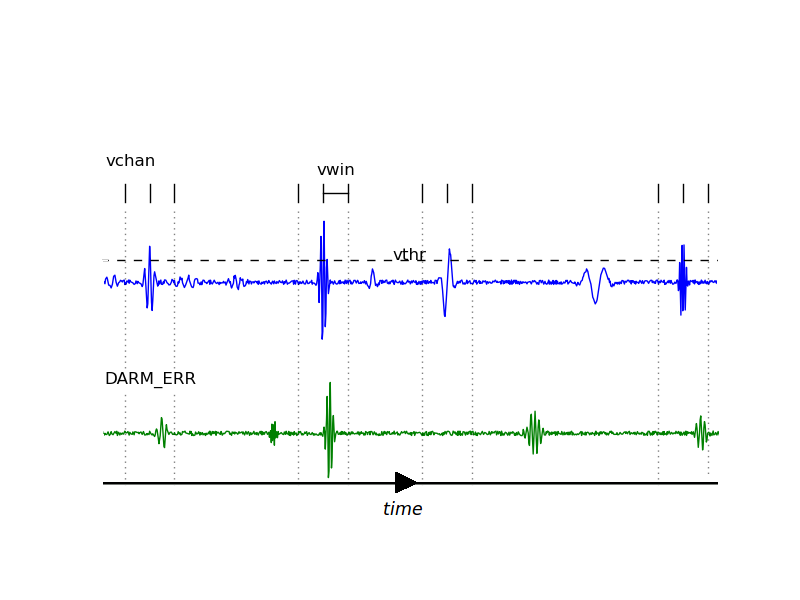}
    \caption{A cartoon showing how veto segments are constructed for a given configuration. Time runs on the x-axis while the y-axis is an arbitrary quantity that may represent the amplitude of the signal in the time-domain or some scalar quantity reflecting its significance in a time-frequency decomposition. In an auxiliary channel (labeled ``vchan'' here), a threshold is applied to the glitches (shown as ``vthr'') and then windows (indicated as ``vwin'' above) are created surrounding those glitches. The union of these windows is then used to remove time from the GW channel (shown as ``DARM\_ERR'' above), thereby vetoing some GW-channel glitches.}
    \label{vconfigCartoon}
  \end{center}
\end{figure}

The OVL algorithm systematically searches for coincident signals in the GW and
auxiliary channels by identifying glitches in the GW channel that fall within
pre-defined time-window surrounding glitches in an auxiliary channel.  In order
to accommodate the variety of possible couplings between auxiliary channels and
the GW channel, the algorithm tries a variety of window sizes and thresholds on
auxiliary glitch thresholds. Each configuration is labeled by a set of three
parameters:  auxiliary channel name, a threshold on the significance of
auxiliary glitches, and time window. 
OVL uses all permutations of auxiliary channels, significance-thresholds and time-windows to create a diverse set of configurations. 
In the specified auxiliary channel, OVL selects only those glitches with significance above the threshold, and then time-windows are constructed around
the central time of each remaining glitch. The union of these time windows
forms a list of (possibly) disjoint segments that are used to remove livetime.
Figure \ref{vconfigCartoon} demonstrates this procedure using some artificial
data.

The algorithm creates these configurations to separate and categorize
different types
of auxiliary glitches. It then applies the configurations to the GW-channel data and
searches for an optimal order. This
allows the method to find highly correlated configurations
while removing spurious information from uncorrelated channels, thereby
identifying troublesome auxiliary channels or glitches. 

There is no reason the parameters describing each configuration must
be limited to channel name, time-window, and significance threshold.
By including other degrees of freedom such as
frequency bands, glitch duration, data epoch, or time of day (e.g. mornings vs.
evenings), the algorithm will be able to find more specific correlations between
configurations and GW-channel data, giving better overall performance. However, the
total number of configurations must also be balanced against the number of
auxiliary glitches available in order to maintain sufficient statistics.

\subsection{Initial application of veto configurations to GW-channel data}\label{Initial application of veto configurations to GW data}

OVL is based on the idea of ordering veto configurations by their
correlations with GW-channel data. We achieve this by associating a figure-of-merit
with each veto configuration. Initially, we assume there is no way to know
which configuration will perform well, and we give each an equal
opportunity{\footnote{Knowledge of the couplings between instrument channels
may provide guidance in defining lists of channels that could make better
or worse vetoes.}}. Each veto configuration is applied to the {\em entire}
GW-channel data set, which we represent as a list of times corresponding to GW-channel glitches. For each veto configuration, veto-segements are
generated as prescribed in Section \ref{Construction of veto configurations}.
We then count the number of GW-channel glitches that fall within these segments,
and determine the configuration's {\em efficiency}, the fraction of events
removed. Likewise, we determine the {\em fractional dead-time} for the
configuration by measuring the fraction of livetime contained in the veto
segments using explicit segment arithmetic. The ratio of these two quantities
defines OVL's figure-of-merit, which we will refer to as efficiency-over-dead-time.
This ``base-line'' efficiency-over-dead-time measurement is then used to
find an initial ordering scheme for all veto configurations.
Figure \ref{vconfigCartoon} shows how this looks when comparing two time series.

We should note that the efficiency-over-dead-time has a straightforward
interpretation in terms of Poisson processes. We can write the efficiency as
$\varepsilon = n_c/N_{GW}$ and fractional dead-time as $f = t / T$ where $n_c$
is the observed number of coincident events, $N_{GW}$ is the total number of GW-channel glitches, $t$ is the amount of time contained in the veto-segments and $T$ is
the total amount of live-time. We then have 
\begin{equation}
\varepsilon / f = \frac{n_c}{t (N_{GW} / T)} \approx \frac{n_c}{t \lambda_{GW}}
\end{equation}
where $\lambda_{GW}$ is an estimate of the rate of GW-channel glitches. This means
that $t \lambda_{GW} \approx \left<n_c\right>$, the expected number of GW-channel
coincidences that will randomly fall within the veto-segments. We then see that
the efficiency-over-dead-time is nothing but the ratio of the observed
coincident events to the expected number of coincident events. As we will see
in the results and analysis section, we observe
efficiency-over-dead-times as high as $10^4$ for some highly-correlated configurations.

Besides associating an efficiency-over-dead-time with each configuration, OVL
also records the Poisson significance of finding the measured number of
coincidences. We compute the Poisson significance as the cumulative probability
of observing as many or more coincident GW-channel glitches given the expected number
of coincident glitches.
OVL uses this significance threshold to avoid over-training by requiring the
observed correlations to be reasonably unlikely to occur by chance given the
large number of configurations tested.

We should note that rankings based on efficiency-over-dead-time and rankings
based on Poisson-significance (as is done with {\em h-veto}) are not equivalent,
and the two metrics may generate different lists. We can see this by noting
that the statistical significance of observing a given number of coincidences
can be written as
\begin{equation}
p = \sum_{k=n_c}^{\infty} \frac{\left<n_c\right>^k}{k!} e^{-\left<n_c\right>} =  \sum_{k=n_c}^{\infty} \frac{\left(n_c(f/\varepsilon)\right)^k}{k!} e^{-n_c(f/\varepsilon)}.
\end{equation}
We then see that the statistical significance is a function of both the number
of coincident events and efficiency-over-dead-time. Generally, for a given
$\varepsilon/f$, the Poisson statistical significance will favor veto
configurations with large $n_c$, and therefore favors configurations with large
number statistics versus a ranking based only on $\varepsilon/f$.

\subsection{Iteration and ordering}\label{Iteration and ordering}

Once a ``base-line'' efficiency-over-deadtime has been determined for each veto configuration, they are ordered accordingly. At this point, veto configurations are applied hierarchically. When a configuration is applied, OVL counts the number of coincident GW-channel glitches and evaluates the associated efficiency-over-deadtime and Poisson-significance. The GW-channel glitches and veto-segments associated with this configuration are removed from the rest of the analysis. In this way, OVL prevents redundant vetoes. Later configurations do not count GW-channel events or live-time already removed by previous configurations. This data-reduction scheme is applied after each configuration, and therefore each veto configuration in the list sees a slightly different set of GW-channel glitches. Furthermore, the statistics computed for each configuration do not represent global fractions, but rather are fractions of a subset of data. This also allows OVL to measure the additional information contained in later configurations. If they are completely redundant with an earlier configuration, then they will not find any new GW-channel glitches and their efficiency-over-deadtime will vanish. In this way, OVL can remove spurious, extraneous, or redundant veto configurations.

This begs the question of what determines an important configuration. We define
these as configurations that yield both a sufficiently high
efficiency-over-deadtime and Poisson-significance. Because these are
functionally dependent, we can think of this as requiring a sufficient strength
of correlation and a sufficiently large number of observed coincidences. By
including the Poisson-significance threshold, OVL tends to reject extremely low
number statistics with low correlation strength, which makes the ordered
list's performance more robust when applying it to different data sets. If
configurations do not perform above these two thresholds, they are removed from
the list.
This process helps OVL to converge rapidly, and prevents under-performing
channels which will be removed in the next iteration from affecting the
performance of later channels.  For final evaluation of a list's performance
after the ordering process is complete, all channels in the list are applied.

In practice, the entire method is repeated several times, which allows for repeated evaluation and ranking of the configurations. After each iteration, the veto configurations are re-ordered according to their most recent efficiency--over--dead-time measurements. The next iteration re-applied the configurations in this new order. If that order is optimal, it won't change between iterations. After about 2 iterations, the list gains the bulk of its performance, with a nearly monotonic decrease in efficiency-over-dead-time as one reads down the ordered list. The near monotonicity is caused by the finite number of iterations, as well as phenomena analogous to cycles in Markov processes. Once a sufficiently monotonic list is created, further iterations do not have a large impact on performance, but they can help simplify the list by removing unnecessary configurations, and are relatively cheap to calculate because the final lists are comparatively short.

\section{Results and analysis}\label{ResultsAndAnalysis}

An earlier version of this method~\cite{blackburnthesis} was used to reject
transient noise artifacts in searches for unmodeled bursts in LIGO's fifth science
run (S5) and
Virgo's first science run (VSR1) data~\cite{BurstsS5,BurstsS5VSR1}.
In those searches, the method was
able to reject 13--45\% of single-detector glitches and 5--10\% of
coincident background (which is generally weak) at under 1\% dead-time.  In
order to further study the performance and systematics of the OVL
procedure, we analysed the entirety of LIGO S4 data (February 22, 2005 - March
25, 2005) from the LIGO Hanford Observatory (H1) and one week of LIGO S6 data
(May 28, 2010 - June 4, 2010) from the LIGO Livingston Observatory (L1) using
KW glitches~\cite{KW2004}. These two data sets were collected from geographically distant detectors and are separated by several years. They therefore represent very different noise environments. Furthermore, because the LIGO detectors underwent significant commissioning between S4 and S6, these data sets may capture different couplings between auxiliary channels and the GW-channel data. Therefore, the similarities in OVL's performance on these very different data sets allows us to draw conclusions about the method rather than peculiarities in either data set.

In this analysis, because there may be causal relations between the GW-channel and auxiliary channels, we restrict ourselves to a subset of auxiliary channels shown to be minimally coupled to the GW channel and thus safe to be used as vetoes. These safety relations are determined through hardware injections at the sites, in which the test masses are driven with a known GW-like transient waveform. Searches for these transients in auxiliary channels during hardware injections at the instruments have been used to establish the exact conditions on their strength (either absolute or related to the GW-channel signal) that make them ``safe'', i.e., ensuring that they will not systematically reject an astrophysical GW transient~\cite{BurstsS4,BurstsS5,BurstsS6}.

In what follows, we find similar performance between the two
data sets, and describe the characteristics of the method. In both cases, OVL
identifies a subset of channels that appear to be correlated with the GW-channel data
which is significantly smaller than the set of all auxiliary channels.

Our implementation of this algorithm used python. An AMD 2.7 GHz 32-bit processor processed $3.7\times10^4$ seconds of data containing $2.8\times10^3$ GW-channel glitches with 250 auxiliary channels from S6 in approximately 9 hours. Nearly half the computational effort is spent on the first two iterations, with the following iterations processing much faster.

\subsection{ROC curves and bulk statistics}\label{ROC curves and bulk statistics}

\begin{figure}
  \begin{center}
    \includegraphics[width=0.5\textwidth]{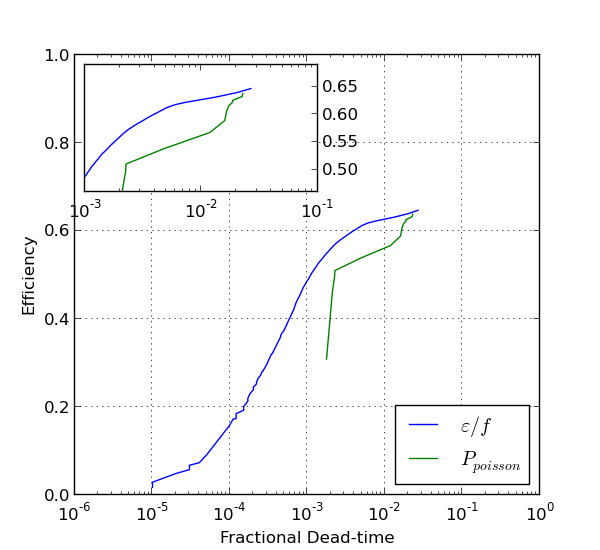}
  \end{center}
  \caption{Comparison of ranking by two figures of merit. Both curves are generated over the S6 data set, using the same code, set of configurations, and performance thresholds. These curves represent the 9$^{\mathrm{th}}$ iteration through OVL, when both ordered lists have settled down to a near optimal order. We see that the efficiency-over-deadtime ($\varepsilon/f$: blue) curve is better than the Poisson Significance curve ($P_{poisson}$: green)). Efficiency--over-dead-time also produces a much smoother curve, indicating that Poisson significance favors larger-number statistics. }
  \label{Psigthr-effbydt}
\end{figure}

\begin{figure}
  \begin{center}
    \begin{minipage}{1.0\linewidth}
      \includegraphics[width = 0.5\textwidth]{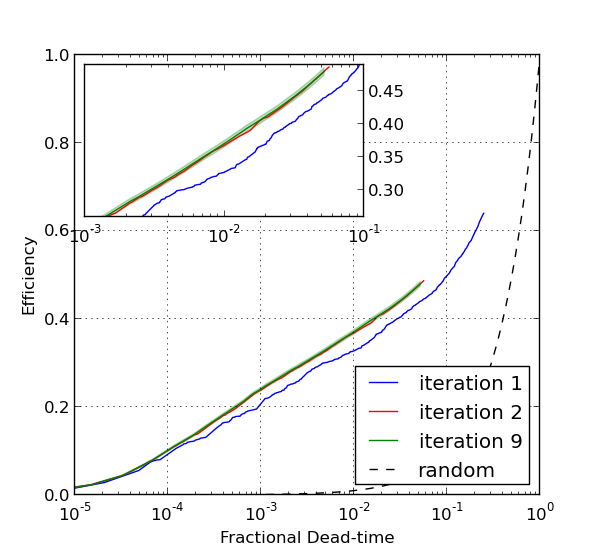}
      \includegraphics[width = 0.5\textwidth]{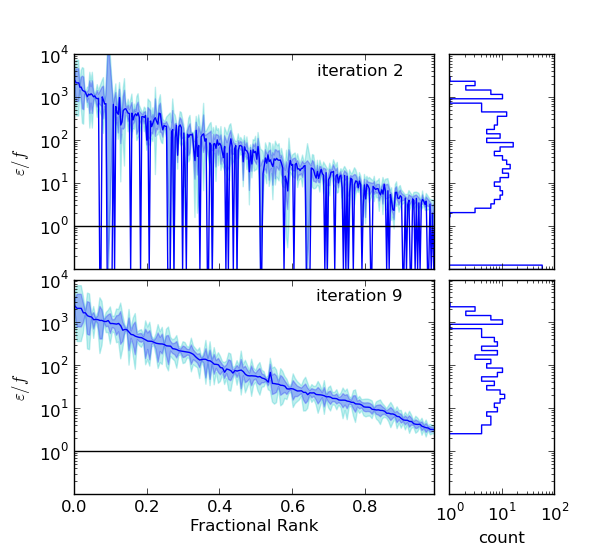}
      \begin{center}
        (a) S4 H1
      \end{center}
    \end{minipage}
    \hspace{0.0cm}
    \begin{minipage}{1.0\linewidth}
      \includegraphics[width = 0.5\textwidth]{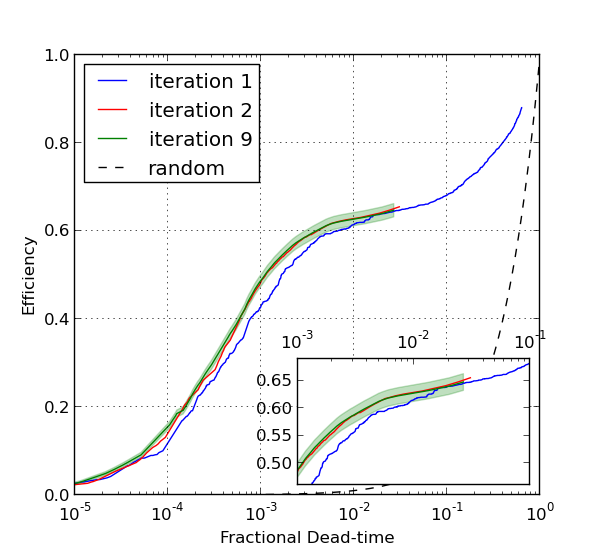}
      \includegraphics[width = 0.5\textwidth]{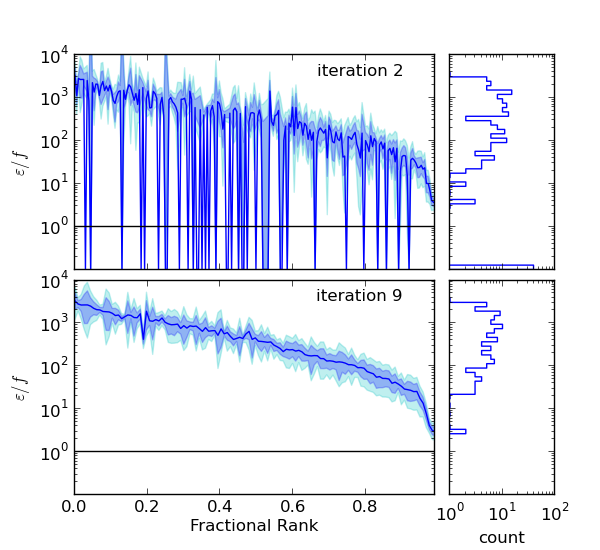}
      \begin{center}
        (b) S6 L1
      \end{center}
    \end{minipage}
  \end{center}
  \caption{(Left) Receiver Operating Characteristic curves. These plots show OVL's performance after different numbers of iterations. Importantly, iteration 1 represents the ordering generated by applying individual veto configurations independently to the entire data set. We would expect the algorithm to improve with further iteration because it can better order its list of vetoes, but we see that the bulk efficiency gains are achieved after only two iterations. There are marginal gains with further iterations, but these are well within the expected errors. The shaded regions represent 68\% confidence intervals. 

(Right) The relation between OVL's figure of merit ($\varepsilon/f =$
efficiency-over-deadtime) and the veto-configuration's order in the list for S4
and S6 data. We expect a nearly monotonic decrease for well-trained lists. The
blue curve denotes the point-estimate, with the shaded regions representing
68\% and 95\% confidence intervals. We see that the errors are relatively small
in these plots, in contrast to Figure \ref{t0wErrors} where the same analysis
is repeated for uncorrelated data. There is the additional
feature in the S6 data: a distinct knee in these plots at high fractional rank.
This suggests a large special population of easily-vetoed glitches.
The projected histograms show the grouping of $\varepsilon/f$, with the peak
clearly above Poisson coincidence's prediction of 1. The bin with $\varepsilon/f=0$ corresponds to configurations that do not find any coincident GW-channel events.}
  \label{ROC}
\end{figure}

The Receiver Operating Characteristic (ROC) curve is the basic diagnostic for
how the OVL algorithm performs. It shows the fraction of GW-channel glitches removed as
a function of the fraction of live-time removed. As a first experiment, we
compared two ranking figures-of-merit: efficiency-over-deadtime and the Poisson
significance, both described in Section \ref{Initial application of veto
    configurations to GW data}. Figure \ref{Psigthr-effbydt} shows the ROC
curves for each ranking scheme. As expected, ranking by incremental
efficiency-over-deadtime results in a higher-performing ROC when measuring
overall cumulative efficiency as a function of dead-time, though the two
rankings begin to converge near at the end where all effective veto conditions
are used up.  Furthermore, the Poisson significance selects many fewer
configurations, each covering a larger number of events. This is seen clearly
by the discreteness of the curve. A short veto list is often convenient for
simplicity and ease of interpretation. A smooth curve with many configurations may
also be preferable in cases where one requires a continuous ranking parameter.

As stated above, OVL's performance improves upon iteration, but the majority of
the efficiency gains are achieved after 2 iterations. Figure \ref{ROC} shows
OVL's performance after several different iterations.  The main advantage of
running to 9 iterations is a reduction in the number of veto configurations
identified.  Table \ref{vconfig_and_vchan} shows that the number of important
channels stabilizes rather early, but the number of important configurations
continues to decrease. This means that further iteration helps to compress the
information stored in the important channels into a shorter list and determine
exactly which glitches in which channels are most troublesome.

\begin{table}[hb]
  \caption{Number of auxiliary channels and veto-configurations present in OVL
as a function of iteration number. We see that the number of important channels
stabilizes quickly and corresponds to the bulk efficiency gains seen in Figure
\ref{ROC}. The information in these channels is then
compressed into a small number of veto-configurations 
(corresponding
to different values of auxiliary channel glitch significance and coincidence
time-window)
upon further iterations.}
  \begin{center}
    \begin{tabular}{@{\extracolsep{\fill}} |c|c||c|c|c|c|c|c|c|c|c|c|}
      \hline
      \multicolumn{2}{|c||}{iteration} & \small{initial} & 1           & 2    & 3   & 4   & 5   & 6   & 7   & 8   & 9 \\
      \hline
      \hline
      \multirow{2}{*}{S4 H1} & No. chan & 161 & 106 & 99   & 52  & 47  & 47  & 47  & 47  & 47  & 47 \\
                 & No. config        & 7245 & 3117 & 294  & 196 & 183 & 178 & 176 & 176 & 176 & 175 \\
      \hline
      \multirow{2}{*}{S6 L1} & No. chan & 202 & 44          & 37   & 35  & 35  & 35  & 35  & 35  & 35  & 35 \\
                 & No. config        & 11250 & 4361      & 209  & 140 & 128 & 119 & 118 & 118 & 117 & 117 \\
      \hline
    \end{tabular}
    \label{vconfig_and_vchan}
  \end{center}
\end{table}

We also note that the ordering improves upon further iteration. Figure \ref{ROC} shows the relation between a configuration's efficiency-over-deadtime and its position in the list for different iterations. We observe a general smoothing with further iterations. Even though the bulk of OVL's performance is gained after 2 iterations, the list still contains many irrelevant and underperforming configurations. By iteration 9, these configurations have been removed and efficiency-over-deadtime decrease much more smoothly, although there are still some fluctuations.  Furthermore, the histograms of efficiency-over-deadtime show the distribution of performance over the list. We see that the peak of the distribution is well away from the Poisson coincidence prediction of 1, and iteration does not damage this distribution.

\subsection{Performance on uncorrelated data}\label{RandomData}

\begin{figure}
  \begin{center}
    \begin{minipage}{1.0\linewidth}
      \includegraphics[width=0.5\textwidth]{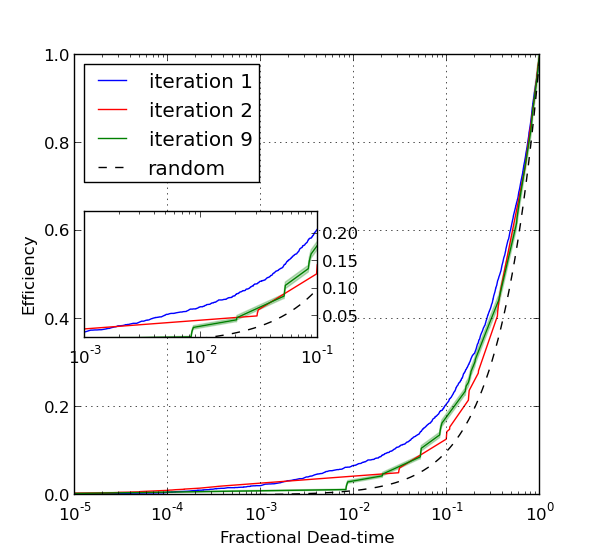}
      \includegraphics[width=0.5\textwidth]{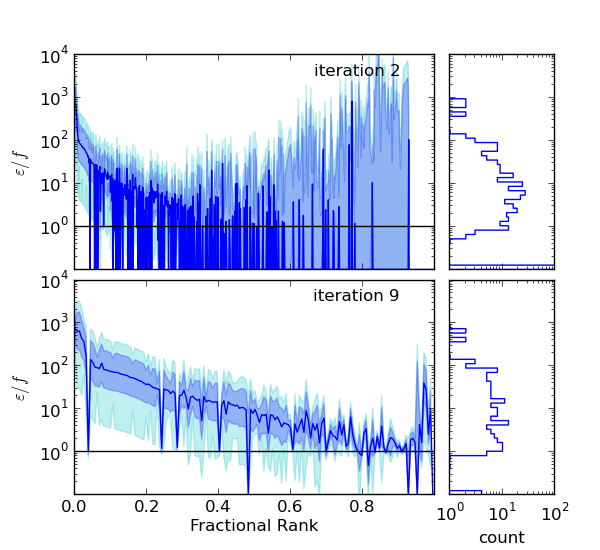}
      \caption{(Left) Receiver Operating Characteristic curves for uncorrelated data. We see that the ROC curves are much worse than those for correlated data. (Right) The relation between OVL's figure of merit and the veto-configuration's order in the list for time-shifted S6 data. The bin with $\varepsilon/f=0$ corresponds to configurations that do not find any coincident GW-channel events. We see that there is still the general trend seen in Figure \ref{ROC}, however the errors are somewhat larger here. This is most noticeable because the upper error estimates are significantly higher in this figure. This is expected, because all coincident GW-channel glitches are due to pure chance in the time-shifted data, and the errors are dominated by small number statistics and the lower bound $n_c \geq 0$.}
      \label{t0wErrors}
    \end{minipage}
  \end{center}
\end{figure}

In order to further check the overall implementation and performance of our
veto algorithm, we examined its output when it was applied on uncorrelated data.
For this purpose we constructed artificial data sets by randomly
shifting the auxiliary glitches in every channel by a different
amount of time using S6 data. In this way, we
break all temporal correlations between the GW channel and auxiliary channels,
as well as between the auxiliary channels themselves. We therefore
expect OVL to be subject only to correlations due to statistical
fluctuations in the coincidence of two Poisson processes. We processed
this data through the OVL pipeline and examined the ROC curve as well as
the dependence of the algorithm's figure of merit as a function of a configuration's fractional rank. This is shown
in Figure \ref{t0wErrors}.

When compared with the plots in Figure \ref{ROC} we see that the ROC curve is {\em much} worse for the uncorrelated data, as expected. However, there are common features in the lists, namely
the comparable decay in efficiency-over-deadtime as one moves to higher
ranks.

Figure \ref{t0wErrors} shows the relation between efficiency-over-deadtime and a configuration's rank. This ordering consists of over-training only because it represents OVL's performance when evaluated on the same data used to generate the ordered list. We should note that when we apply the threshold on Poisson significance used for the correlated data to this uncorrelated data, only $1\%$ of these configuration survive. However, by lowering the Poisson significance threshold, we are able to examine the underlying distribution of noise in our analysis. We can immediately interpret our threshold on efficiency-over-deadtime from the projected histogram as designed to separate the distributions for correlated data (Figure \ref{ROC}) and uncorrelated data (Figure \ref{t0wErrors}).

\subsection{Round robin algorithm}\label{Round robin algorithm}

\begin{figure}
  \begin{minipage}{0.5\linewidth}
    \includegraphics[width=\textwidth]{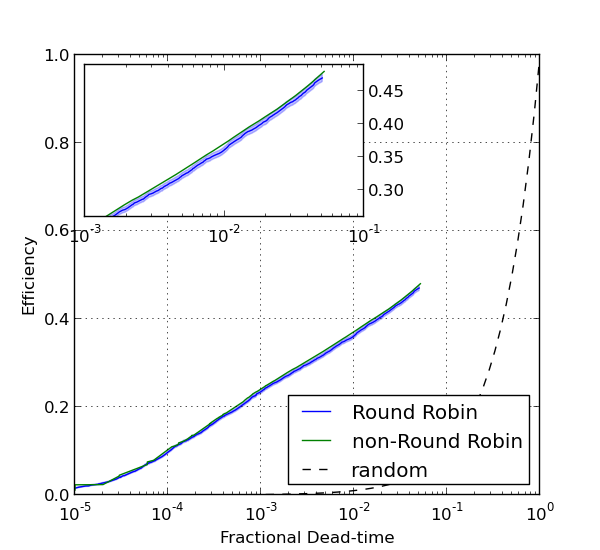}
    \begin{center}
      (a) S4 H1
    \end{center}
  \end{minipage}
  \begin{minipage}{0.5\linewidth}
    \includegraphics[width=\textwidth]{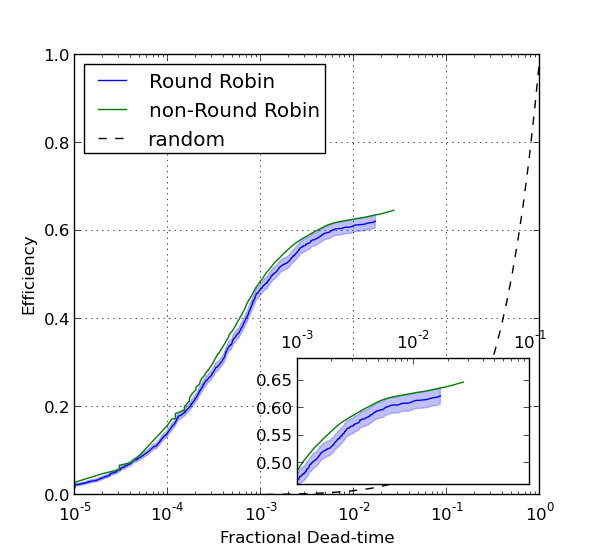}
    \begin{center}
      (b) S6 L1
    \end{center}
  \end{minipage}
  \caption{A comparison of round robin and non-round robin analysis techniques.
      Shaded regions represent 68\% confidence intervals. The round robin
      algorithm ensures that lists are developed and then evaluated using
      disjoint data sets. However, we see that the application to disjoint data
      has a very small affect on the list's performance. This suggests that OVL
      finds overwhelmingly true correlations in the data. The observed decrease
      in performance could be due to a combination small over-training errors
      as well as statistical fluctuations caused by evaluating a list's
      performance on a smaller data set (implicit in the division associated
      with round robin analysis).
}
  \label{roundRobin}
\end{figure}

There exists a potential danger of over-training OVL with a single data set,
which may produce artificially high performance that will not generalize to
other data. For this purpose, we performed a simple cross-referencing
procedure that shows this is not significant for our results. We divided the
data into separate sets, which were arranged like a 1-dimensional checker-board in time, meaning the $k^\mathrm{th}$ bin contains every $k^\mathrm{th}$ minute of the livetime.
Training was then performed on all sets except one, and that one set was used
for evaluation. This was carried out with each set, so we have a measure of the
algorithm's performance over all our data.  Fig \ref{roundRobin} shows a
comparison between the round robin and non-round robin procedure. We see that
there is no significant difference between the two curves at the 68\%
confidence level, although there is a small systematic error introduced. The
consistency between round robin and non-round robin performance is due to OVL's
threshold on Poisson significance, described in Section \ref{Initial
    application of veto configurations to GW data}. By requiring the number of
observed coincidences to be statistically significant relative to the total
number of veto configurations tested, OVL can reject configurations with low
number statistics that may pass the performance threshold due to random
coincidence alone.

\begin{figure}
  \begin{center}
    \includegraphics[width=0.5\textwidth]{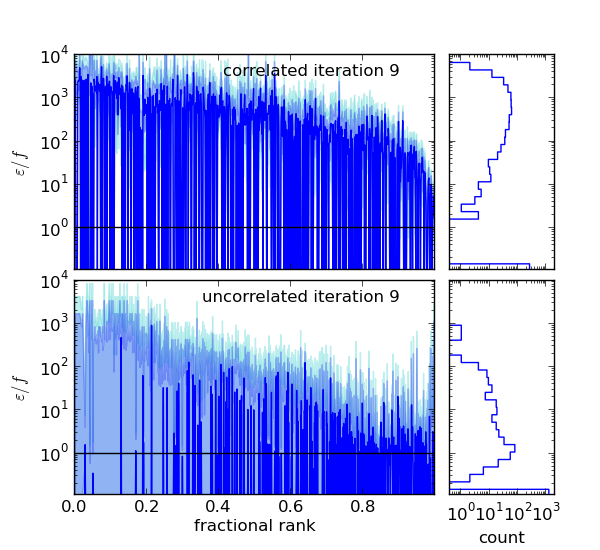}
    \caption{Comparison of OVL figure-of-merit upon evaluation using round-robin binning and correlated/uncorrelated data. We see that there are many configurations that perform poorly upon round-robin evaluation in both the correlated and uncorrelated data. However, we also see that the histograms of the two data sets are much different, with clearly separated points of maximum likelihood. Furthermore, the uncorrelated data appears to have a peak in likelihood near $\varepsilon/f\approx1$, which is exactly what we expect from Poisson coincidence. The bin with $\varepsilon/f=0$ corresponds to configurations that do not find any conincident GW-channel events.}
    \label{S6-rr-effbydt}
  \end{center}
\end{figure}

We also examine the performance of individual configurations when round robin binning is used. Figure \ref{S6-rr-effbydt} shows the efficiency-over-deadtime for correlated and uncorrelated S6 data (discussed further in Section \ref{RandomData}), and should be compared with Figures \ref{ROC} and \ref{t0wErrors}. It is clear that both correlated and uncorrelated data contain some configurations that do not perform well on disjoint data. These are likely due to statistical fluctuations and represent the sharp spikes in these plots. However, we also see that consistent performance is the norm in correlated data, whereas under-performance is the norm for uncorrelated data. The few outliers in the uncorrelated data set are due to single auxiliary glitches that happen to coincide with GW-channel events, and are statistical in nature. We also note that the distributions over efficiency-over-deadtime are very different for the two data sets. Encouragingly, the uncorrelated data appears to have a peak in likelihood near $\varepsilon/f=1$, which agrees with our Poissonian interpretation. We should also note that the ordering in the uncorrelated data is generated entirely by statistical errors in the estimation of efficiency and dead-time, which is clearly visible in the error estimates for uncorrelated data show in Figure \ref{S6-rr-effbydt}.

\subsection{Effects of algorithmic parameters}\label{Effects on algorithmic parameters}

OVL constructs a set of configurations based on a list of channels,
thresholds, and time-windows, using all possible permutations of these parameters. In the bulk of this paper, we used time-windows from the set \{25ms, 50ms, 100ms, 150ms, 200ms\} and KW significance thresholds from the set \{15, 25, 30, 50, 100, 200, 400, 800, 1600\}. These thresholds and windows were chosen because they reflect the types of glitch significance and correlation time-scales observed in the past. We would expect that increasing the dimensionality
of these configurations would allow the algorithm to better separate classes
of glitches (characterized by the elements of the configuration).

Generally, we see that high auxiliary significance thresholds congregate at the top of the ordered list. This is because these events are relatively rare and are likely to be strongly correlated with GW-channel glitches. We contrast this with low threshold events, which are relatively common and may be more strongly influenced by a statistical background. This also means that these high-threshold auxiliary glitches are used first and dominate OVL's performance at low dead-time. However, at sufficiently high dead-time, low-significance auxiliary glitches begin to dominate performance. Consider a set of veto configurations with the same auxiliary channel and significance threshold. Then, for a configuration from this set to remove more GW-channel glitches, it must widen the time-window applied around auxiliary glitches. If the GW-channel glitches are clustered, this can improve efficiency, but if the GW-channel glitches are sparse, it will mainly increase fractional dead-time without removing more glitches. However, lower threshold events may be able to better select small, widely separated GW-channel glitches without increasing the time-window used. Even if the correlation is weaker, this can increase overall performance. Similar effects are observed with the time-windows. Small time-windows typically show up at the top of the list with larger time windows appearing later and at decreased efficiency-over-deadtime.

\section{Applications in instrument characterization}\label{DetChar}

The OVL method we just described generates a wealth of information beyond
the list of times that should be used as vetoes in a search for
gravitational-wave transients. This information can be mined and used for
general instrument monitoring and characterization. The algorithm is also
simple and efficient enough so that to be able to analyze data within
a few seconds from real time and without any significant impact on
computational resources.

In giving a flavor of the capacity of our analysis, we will focus on the
immediately derived quantities out of our statistical analysis\footnote{The
    individual raw glitches used as input as well as the time series of vetoes
    produced can also be mined for even richer statistical information for
    instrument characterization. This includes analysis on Poissonianity,
    frequency content, and the specific role they may play on glitches in the
    GW channel of certain strength, frequency or waveform morphology.}.  The
typical output of our algorithm in the training phase is captured in Figure
\ref{OVLoutput}, where the most relevant auxiliary channels (and their
corresponding parameters) are ranked using our unified criterion.

\begin{table}[hb]
  \caption{This is an example OVL output from iteration 9 of the S6 data. We observe a nearly monotonic decrease in efficiency-over-deadtime as we proceed down the list. We also notice that several channels appear multiple times with different sets of significance thresholds and time-windows. In way of definition, \emph{livetime} is the number of seconds in the analysis; \emph{\#gwtrg} is the number of GW-channel glitches in the analysis; \emph{vchan} is the auxiliary channel name; \emph{vthr} is the threshold on auxiliary KW significance; \emph{vwin} is the time-window applied around each auxiliary glitch; \emph{deadsec} is the number of seconds removed by this configuration; \emph{vexp} is the expected number of GW-channel glitches coincident contained in \emph{deadsec}; \emph{vact} is the actual number of coincident GW-channel glitches; \emph{vsignif} is the negative-logarithm of the Poisson significance of observing \emph{vact} coincident GW-channel glitches when we expected \emph{vexp}; \emph{eff} is the efficiency; \emph{fdt} is the fractional dead-time; \emph{eff/fdt} is efficiency-over-deadtime, OVL's ranking metric.}
  \begin{center}
{\scriptsize
\begin{verbatim}
  livetime #gwtrg  vchn                               vthr   vwin deadsec  vexp vact vsignif  eff/fdt 
374014.000   2826  L1_LSC-POB_Q_1024_4096              400  0.025   0.238  0.00    6   44.50  3334.29
374013.762   2820  L1_OMC-PZT_LSC_OUT_DAQ_8_1024      1600  0.025   1.500  0.01   32  225.00  2829.42 
374012.262   2788  L1_OMC-PZT_LSC_OUT_DAQ_8_1024       400  0.025   0.550  0.00   11   77.97  2683.02 
374011.712   2777  L1_ISI-OMC_GEOPF_H2_IN1_DAQ_8_1024 1600  0.100   0.150  0.00    3   22.19  2693.64 
374011.562   2774  L0_PEM-LVEA_SEISZ_8_128             200  0.050   0.100  0.00    2   15.11  2696.55 
374011.462   2772  L1_ISI-OMC_GEOPF_H1_IN1_DAQ_8_1024  400  0.025   0.266  0.00    5   35.94  2539.06 
374011.196   2767  L1_OMC-PZT_LSC_OUT_DAQ_8_1024       200  0.025   0.541  0.00    9   62.50  2250.26 
374010.656   2758  L1_OMC-PZT_LSC_OUT_DAQ_8_1024       100  0.025   0.908  0.01   14   95.28  2090.54 
374009.747   2744  L1_ASC-ITMY_P_8_256                 400  0.025   3.700  0.03   56  374.35  2062.93
374006.047   2688  L0_PEM-LVEA_BAYMIC_8_1024           400  0.025   1.835  0.01   25  166.23  1895.94
\end{verbatim}}
    \label{OVLoutput}
  \end{center}
\end{table}

\begin{figure}
  \begin{center}
    \includegraphics[height=0.9\textheight]{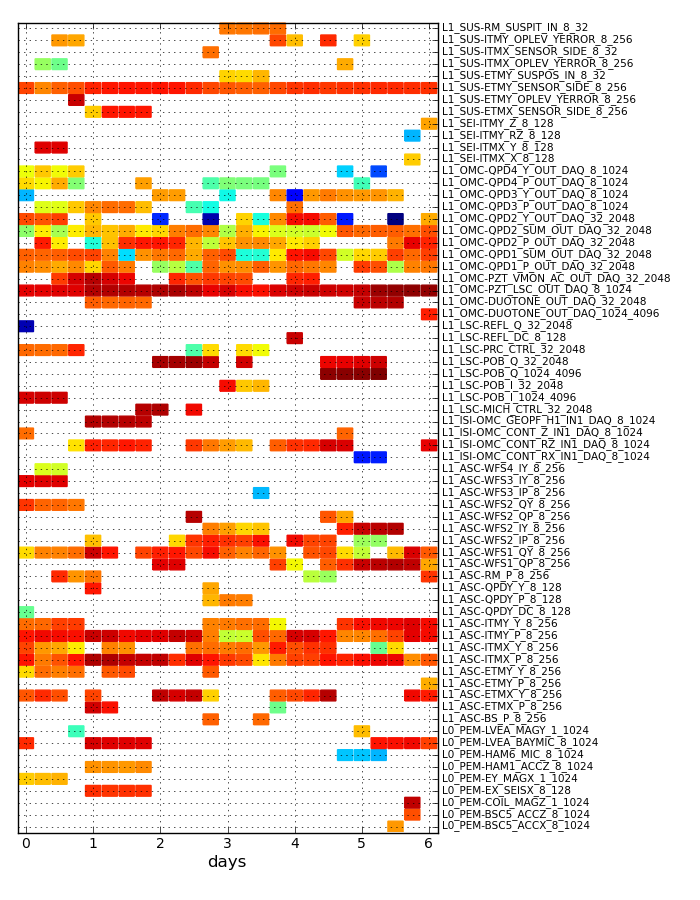}
    \caption{Appearance of auxiliary channels shown over the single week of S6 data. Channels' relative performance is encoded in the color scheme, with dark red implying high $\varepsilon/f$ and blue low $\varepsilon/f$. Simple thresholding on absolute significance or relative changes in significance can provide handles for initiating and assisting investigations on instrument performance. We should note that the ``L1'' prefix on the channel names corresponds to the LIGO Livingston detector, and the last two numbers at the end of the channel name correspond to the bandwidth of that channel (in Hertz).}
    \label{ChannelsOverTime}
  \end{center}
\end{figure}

The identification and ranking of channels that contribute to
gravitational-wave-like glitches allows tracking such
contributions over time, thus promptly signaling changes in the couplings between
instrument channels and their environments. In Figure \ref{ChannelsOverTime}
we show an example of how auxiliary channels appear in OVL's final 
training table (like the one in Figure \ref{OVLoutput})
The colors reflect the significance of each channel in the
corresponding veto configuration list, with dark red corresponding to high efficiency--over--dead-time and blue corresponding to low efficiency--over--dead-time. Applying simple thresholds either
in the absolute significance or in the stride-by-stride changes in significance
can provide useful handles in identifying and investigating changes
in instrument performance.

\begin{figure}
  \begin{center}
    \includegraphics[width=0.75\textwidth]{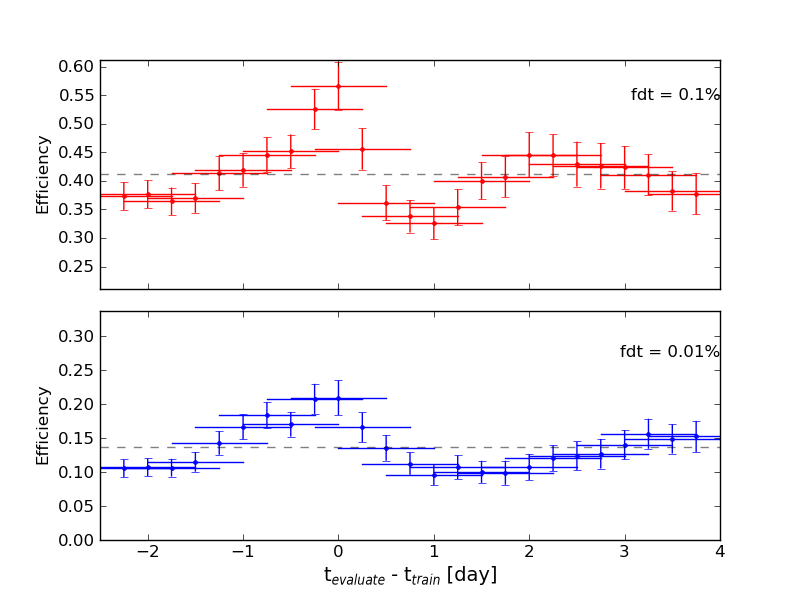}
    \caption{Efficiency at fixed fractional dead-time over 1 week of S6 data. This plot shows the change in performance for a single ordered list when it is applied on different days. The abscissa shows the time difference between the training and evaluation sets and we see that there is significane flucutation in performance, although those fluctuations appear to be bounded.}
    \label{effdtplot}
  \end{center}
\end{figure}

Zooming out to more macroscopic quantities, key summary quantities encoded
in our ROC curves like efficiency and dead-time can be readily used for 
prompt figure-of-merit criteria when making data acquisition decisions as
data come in. For example, by fixing a tolerable dead-time one can
monitor the efficiency of the veto algorithm as a function of time, or in
a symmetric way monitor the dead-time corresponding to a veto configuration that
results in a fixed efficiency. We show one such example in
Figure \ref{effdtplot}. In this plot the resulting efficiency in removing
gravitational-wave-like glitches from the gravitational-wave channel
over seven days of S6 data is shown at two arbitrary dead-times, one at
0.1\% and another one at 0.01\%.

The examples we have used above mostly target identification and studies
of what has been a commonly encountered feature of the first generation
gravitational-wave detectors, namely, non-stationarity. Our algorithm
assists the study of such variability at the single-channel level
as well as at the macroscopic level, in terms of quantities that will end
up being relevant in an end-to-end search (like residual singles rates
before and after veto application).

We should also note that OVL can be used to conduct safety studies, in which we determine which auxiliary channels are coupled to possible GW signals. The results presented in this paper are predicated on previous safety studies, conducted via hardware injections. However, OVL need not resort to these external studies. By running the OVL algorithm over a set of fake glitches corresponding to \emph{only} hardware injections, OVL will naturally select the auxiliary channels that are highly correlated with these injections. By removing these auxiliary channels and re-processing the hardware injections, we can iteratively determine which auxiliary channels are unsafe and remove them from the analysis. Clearly, the procedure should terminate when OVL's performance becomes consistent with uncorrelated data, and example of which is shown in Section \ref{RandomData}. These safety studies can be repeated for various types of injections in order to determine which auxiliary channels are unsafe for certain classes of expected waveforms.

\section{Conclusions}\label{Conclusions}

A variety of astrophysical sources of GWs are expected to produce
short-duration signals within the sensitive frequency band of kilometer-scale
ground-based interferometers, such as those operated by the LIGO, GEO600 and Virgo
collaborations.  Multiple terrestrial noise sources can produce
transient artifacts (glitches) that resemble these short duration signals, and
their contribution to the background for searches of these types of signals is
a limiting factor for search sensitivity. An active area of research
involves developing procedures to remove this transient noise during analysis
of GW-channel data by using information derived from the many auxiliary channels which
measure the local environment and other instrumental non-GW degrees of freedom.
We presented the Ordered Veto List (OVL) algorithm, which aims to identify the unique
and most relevant correlations between GW-channel glitches and similar signals
seen in auxiliary channels by use of an iterative application and ranking of
auxiliary channel veto conditions.

OVL uses many veto configurations to describe the types of couplings between
auxiliary channels and the GW channel. This implementation used auxiliary
channel name, a threshold on auxiliary glitch significance, and a time-window
surrounding auxiliary glitches. However, the algorithm can be easily extended
to include other parameters such as frequency information or time of day.
OVL then constructs a list of segments using these configuration parameters.
The union of time-windows surrounding auxiliary glitches with a sufficiently
high significance is used to remove data from the GW channel. OVL then computes
the configuration's \emph{efficiency} and \emph{dead-time}, the fractions of
GW-channel glitches and livetime removed, respectively. The configurations are then
ranked according to the ratio of these numbers, \emph{efficiency-over-deadtime},
which has the natural interpretation as the ratio of the number of removed GW-channel
glitches and the expected number of removed GW-channel glitches. Furthermore, these
configurations are applied hierarchically, in that once a configuration is
applied, later configurations do not see any data removed by earlier configurations.
Upon iteration, this repeated ranking and application finds highly correlated
auxiliary channels and constructs an ordered list based on the
\emph{efficiency-over-deadtimes} observed. It also removes any redundant or
uninformative channels.

This paper presented OVL results based on the entirety of LIGO S4 data
(February 22, 2005 -- March 25, 2005) from the LIGO Hanford Observatory (H1)
and one week of LIGO S6 data (May 28, 2010 -- June 4, 2010) from the
LIGO Livingston Observatory (L1) using Kleine Welle glitches. These two data sets
represent very different noise sources and couplings between auxiliary channels
and the GW channel. Therefore, the common trends we observe in the two data
sets are indicative of OVL itself, rather than any peculiarities within a data set. 

We examined OVL's performance using standard Receiver Operating Characteristic
(ROC) curves, which shows the fraction of noise signals we are able to remove
as a function of the fractional loss in live-time incurred. We see that the
bulk performance is gained after 2 iterations, and this improves over the
application of configurations based solely on their individual performance when
applied to the entier data set. Further iteration helps to compress the useful
information in the ordered list into a small number of configurations. This
helps isolate exactly which correlations are significant and makes the list
easier to read.  We find that while the ROC curve only marginally improves
between 2 and 9 iterations, the ordered list is only 1-2\% as long after 9
iterations.

Finally, we point out a few possible applications of OVL to instrument
characterization, including tracking and accounting for instrument
non-stationarity and identification of subsystems containing channels that
are highly correlated with GW-channel glitches.

\section*{Acknowledgements}
RE and EK gratefully acknowledge the support of the National Science Foundation and the LIGO
Laboratory. LIGO was constructed by the California Institute of Technology and
Massachusetts Institute of Technology with funding from
the National Science Foundation and operates under cooperative
agreement PHY-0757058. 
LB is supported by an appointment to the NASA Postdoctoral Program at
Goddard Space Flight Center,
administered by Oak Ridge Associated Universities through a contract with NASA.
The authors would also like to acknowledge comments and feedback received by members 
of the Bursts, Compact Binary Coalescences and Detector Characterization working groups
of the LIGO Scientific Collaboration and the Virgo Collaboration, as well as the entire LIGO Scientific Collaboration and Virgo Collaboration for access to this data.
This paper has been assigned LIGO Document Number LIGO-P1300038.


\section*{References}
\bibliography{refs}

\end{document}